\documentclass[prl,nofootinbib,twocolumn,showpacs,floatfix]{revtex4}

\usepackage{graphicx}
\usepackage{amsmath, amssymb, amsfonts}

\begin{document}
\title{Energy-gap dynamics of superconducting NbN thin films studied by time-resolved
terahertz spectroscopy}

\author{M. Beck,$^{1}$ M. Klammer,$^{1}$ S. Lang,$^{1}$ P. Leiderer,$^{1}$ V.V.
Kabanov,$^{2,3}$ G. N. Gol'tsman,$^{4}$ and J. Demsar$^{1,2,3}$}
\affiliation{$^{1}$Dept. of Physics and Center for Applied Photonics, Univ. of Konstanz,
D-78457, Germany}
\affiliation{$^{2}$Zukunftskolleg, Univ. of Konstanz, D-78457, Germany}
\affiliation{$^{3}$Complex Matter Dept., Jozef Stefan Institute, SI-1000, Slovenia }
\affiliation{$^{4}$Moscow State Pedagogical University, Moscow, Russia}


\begin{abstract}
Using time-domain Terahertz spectroscopy we performed direct studies of the
photoinduced suppression and recovery of the superconducting gap in a
conventional BCS superconductor NbN. Both processes are found to be strongly
temperature and excitation density dependent. The analysis of the data with
the established phenomenological Rothwarf-Taylor model enabled us to determine
the bare quasiparticle recombination rate, the Cooper pair-breaking rate and
the electron-phonon coupling constant, $\lambda=1.1\pm0.1,$ which is in
excellent agreement with theoretical estimates.

\end{abstract}
\maketitle

Soon after the first tunneling experiments in superconductors (SCs) revealed that
the quasiparticle (QP) tunneling can be qualitatively described in terms of
the QP energy band picture, in which the SC can be treated
similarly to a narrow gap semiconductor, it was realized that SCs can be used
as detectors for far-infrared light \cite{Burnstein}. Since real-world
applications of SCs require an understanding of their properties under
non-equilibrium conditions several attempts were made already in 1960's to
determine the time scales and processes that govern the recovery of the SC
state. It was soon realized \cite{Ginsberg} that the recombination process,
where two QP recombine to form a Cooper pair, was dominated by the emission of
phonons with $\hbar\omega>2\Delta$, $\Delta$ being the SC gap. However, as
pointed out by Rothwarf and Taylor \cite{RothwarfTaylor}, the re-absorbtion of
$\hbar\omega>2\Delta$ phonons leads to the so called phonon bottleneck, where
the recovery of the SC state is not governed by the bare recombination of two
QPs into the condensate, but rather by the decay of $\hbar\omega>2\Delta$ phonons.
The non-equilibrium state can be generated either by QP injection in tunnel
junctions \cite{Ginsberg}, or by excitation with photons of energy larger than
2$\Delta$. In particular, due to the rapid development in the generation of
ultrashort laser pulses and related femtosecond pump-probe techniques, the
latter approach has enabled direct studies of carrier relaxation dynamics with
femtosecond time resolution. Most of the work reported to date focussed on
high-T$_{c}$ SCs; predominantly on cuprates
\cite{Han,Kabanov,mercury,THzAveritt,Gedik,KaindlBiSCO,Perfetti,Kusar,Giannetti,Pashkin,Uwe,Beyer}
but more recently also on pnictides \cite{Mertelj,Chia,Mansart,GedikPnic}. The
main focus of the research was on the influence of the superconducting gap and
the normal state pseudogap on the carrier relaxation dynamics
\cite{Han,Kabanov,mercury,Gedik,KaindlBiSCO,MgB2,RT,nicol,Hirschfeld}. One of
the still controversial issues with cuprates is the effect of the d-wave gap
on the relaxation phenomena, and the role of phonons, i.e. the existence
\cite{Kabanov,RT,Uwe} or absence \cite{KaindlBiSCO,Gedik} of the phonon
bottleneck in this class of SCs. Recently, systematic studies of the
photoinduced melting of SC have been performed in cuprates
\cite{Kusar,Giannetti,Pashkin,Beyer}. Moreover, manipulation of the order
parameter in La$_{1.84}$Sr$_{0.16}$CuO$_{4}$ by applying intense Terahertz
(THz) electric fields along the c-axis has also been demonstrated
\cite{Dienst}. Despite numerous studies in high-T$_{c}$ SCs, whose ground
state properties are not well understood, only a few experiments on
conventional BCS SCs exist \cite{Federici,Carr,MgB2,Lobo}, but no systematic
study of dynamics as a function of temperature (T) and the absorbed energy
density (\emph{A}). Studies of non-equilibrium dynamics in conventional SCs
are an important first step to understanding relaxation dynamics in more
exotic SCs. Since for conventional BCS SC $\Delta$ lies in the low-THz range,
and can be resonantly probed by time domain THz spectroscopy (TDTS), the data
interpretation should be less ambiguous.

In this Letter we report on the first detailed study of the SC state
relaxation phenomena in a conventional SC NbN \cite{Geibel} over wide range of
T and \emph{A}. Utilizing the TDTS we have studied the T-dependence of its
complex conductivity, $\sigma(\omega)$, as well as its T- and \emph{A}%
-dependent dynamics following photoexcitation with a fs optical pulse. We show
that $\sigma(\omega)$ can be well fit to the prediction of the BCS theory
\cite{Zimmermann}, enabling direct studies of the temporal evolution of
$\Delta$. Both, the Cooper pair breaking (CPB) and the SC state recovery
were found to be T- and \emph{A}-dependent, and could be well
explained by the Rothwarf-Taylor (RT) phonon-bottleneck model
\cite{RothwarfTaylor,RT}. From the dependence of the CPB on \emph{A
} we were able to determine the microscopic QP recombination rate, $R$, as well
as the value of the electron-phonon (e-ph) coupling constant, $\lambda$.

NbN thin films were deposited by dc magnetron sputtering on MgO
substrates \cite{Goltsman}. Films with thicknesses between 10-15 nm and
T$_{c}$ between 14.3-15.4 K ($\Delta T_{c}=0.19-0.17$ K) were investigated.
The TDTS set-up was built around a 250 kHz amplified Ti:sapphire laser system,
utilizing large area interdigitated photoconductive emitter for the generation
of THz pulses \cite{Emitter}, while the THz electric fields transmitted
through the sample, $E_{tr}(t^{\prime})$, and reference, $E_{re}(t^{\prime})$,
were detected using the Pockels effect in GaP. The real, $\sigma_{1}(\omega)$,
and the imaginary, $\sigma_{2}(\omega)$, parts of the optical conductivity
were determined using the appropriate Fresnel equations. In non-equilibrium
experiments, the films were excited by 50 fs pulses at the carrier wavelength
of 800 nm. The time evolution of the SC state as a function of $t_{d}$, the
time delay of the THz pulse with respect to the optical excitation, was
studied by either directly measuring $\sigma
(\omega,t_{d})$ or by measuring the induced changes in the transmitted
electric field, $\Delta E_{tr}(t^{\prime},t_{d})$, at a fixed $t^{\prime}$.
The latter, spectrally integrated studies, are particularly useful for
studying dynamics at low excitations \cite{THzAveritt,MgB2,KaindlBiSCO}. The
THz beam diameter on the sample was $\approx1.5$ mm\cite{diameter}, while that
of the photoexcitation beam was 4 mm, ensuring the homogeneous lateral
excitation profile. From reflectivity and transmission of the films at 800 nm
we determined the optical penetration depth $l_{opt}\approx12$ nm. The
absorbed energy densities at the film surface, \emph{A}, were calculated from
the reflectivity and $l_{opt}$ and are given in mJ/cm$^{3}$.%

\begin{figure}
[ht]
\begin{center}
\includegraphics[
height=6.0539cm,
width=8.5009cm
]%
{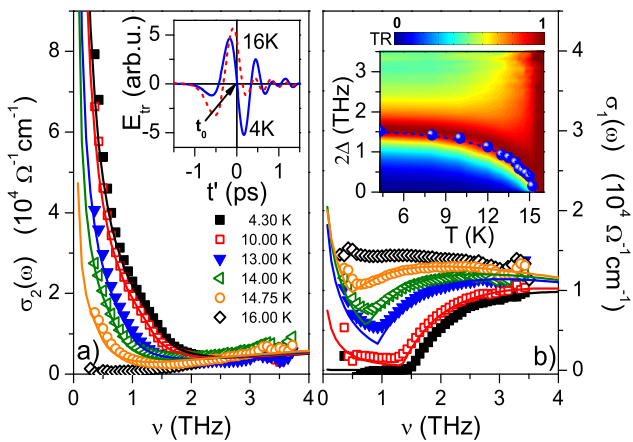}%
\caption{(color online) The T-dependence of the a) imaginary and b) real part
of $\sigma\left(  \omega\right)  $ of 15 nm NbN film on MgO substrate
(T$_{c}=15.4$ K). Solid lines are fits with the BCS equations
\cite{Zimmermann}. Inset to a) shows the transmitted THz transients through
the NbN film below and above T$_{c}$; the arrow denotes the time $t^{\prime
}=t_{0}$ with maximum change in the transmitted electric field. Inset to b):
the T-dependence of $\Delta$, extracted from fits to $\sigma\left(
\omega\right)  $ (symbols) overlaying the normalized transmissivity ratio, TR.
The BCS T-dependence of $\Delta$ is shown by the dashed line.}%
\end{center}
\end{figure}

Fig. 1 presents the T-dependence of $\sigma(\omega)$ of a 15 nm NbN film with $T_{c}=15.4$ K. In the normal state
$\sigma(\omega)$ is fit with the Drude model with the plasma frequency
$\nu_{p}=460$ THz (15300 cm$^{-1}$) and the scattering rate $\hbar/\tau=264$
cm$^{-1}$, in good agreement with studies on thick films \cite{vanderMarel}.
In the SC state an excellent agreement between the data and $\sigma(\omega)$
using the BCS model with a finite normal state scattering rate
\cite{Zimmermann} (solid lines) is obtained. The extracted T-dependence of
$\Delta$, with $\Delta(0)\approx0.75$ THz (3.07 meV) and $2\Delta/k_{B}%
T_{c}\approx4.6$, is shown in inset to panel b). $\Delta(T)$ is plotted on top
of TR, the transmissivity ratio $\mathcal{T(}\omega,T\mathcal{)}%
$/$\mathcal{T(}\omega,16$K$\mathcal{)}$, which is at each T normalized to its
peak value (in a BCS SC, the transmissivity ratio peaks just above 2$\Delta$ \cite{vanderMarel}).

\begin{figure}
[ht]
\begin{center}
\includegraphics[
height=6.3065cm,
width=8.5009cm
]%
{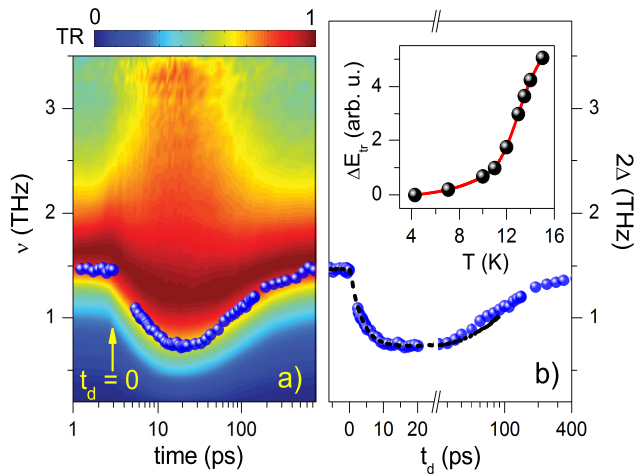}%
\caption{(color online) a) The time evolution of normalized transmissivity
ratio (TR), recorded at $A$ = 22 $m$J/cm$^{3}$, together with $\Delta\left(
t\right)  $ extracted from $\sigma_{1}(\omega,t_{d})$ fit with the BCS formula
\cite{Zimmermann}. b) Comparison of $\Delta(t_{d})$ extracted from $\sigma
_{1}(\omega,t_{d})$ (symbols) and from the spectrally integrated response
$\Delta E_{tr}(t_{0},t_{d})$ (dashed line). Inset: the measured T-dependence
of $\Delta E_{tr}(t_{0})$ used to relate $\Delta E_{tr}(t_{0},t_{d})$ to the
change in $\Delta$. }%
\end{center}
\end{figure}

Fig. 2a) presents the temporal evolution of TR following photoexcitation with
a 50 fs optical pulse with \emph{A} = 22 mJ/cm$^{3}$. The data reveal a strong
suppression of SC on the 10 ps time scale followed by the recovery on the 100
ps timescale. On the same plot $\Delta\left(  t_{d}\right)  $, obtained by
best fit to $\sigma_{1}(\omega,t_{d})$ using the BCS model \cite{Zimmermann}
is plotted by solid symbols.

To perform systematic studies of the time evolution of $\Delta$ as a function
of \emph{A} and T, avoiding the signal drifts associated with the long term
laser stability, we studied the dynamics of the induced changes in the
transmitted electric field, $\Delta E_{tr}(t^{\prime})$, at a fixed point of
$E_{tr}(t^{\prime}=t_{0})$. The fixed point of $E_{tr}(t_{0})$ in these
experiments was at the point of maximum time derivative of the electric field,
$t_{0}=0$ ps, where the changes in $E_{tr}(t_{0})$ upon entering the SC state
are the largest $-$ see inset to Fig. 1a). In order to obtain a direct link
between $\Delta E_{tr}(t_{0})$ and the photoinduced change of the gap,
$\delta\Delta$, we first measured $\Delta E_{tr}(t_{0},T,T_{0})=E_{tr}%
(t_{0},T)-E_{tr}(t_{0},T_{0})$, where $T_{0}$ is the T of the sample at which
photoinduced studies are performed. Combining $\Delta E_{tr}(t_{0},T,4.3$K$)$,
shown in inset to Fig. 2b), with the T-dependence of the gap (inset to Fig.
1b)) enabled us to extract the temporal evolution of the gap from $\Delta
E_{tr}(t_{0},t_{d})$ traces. The trace obtained this way at \emph{A} = 22
mJ/cm$^{3}$, shown by the dashed curve in Fig. 2b), matches well the time
evolution of the gap extracted from the $\sigma_{1}(\omega,t_{d})$ data.
Therefore, by measuring the spectrally integrated response, the T- and
\emph{A}-dependence of the photoinduced gap change, $\delta\Delta$, can be
extracted quickly, avoiding system drifts.%

\begin{figure}
[ht]
\begin{center}
\includegraphics[
height=5.7288cm,
width=8.4987cm
]%
{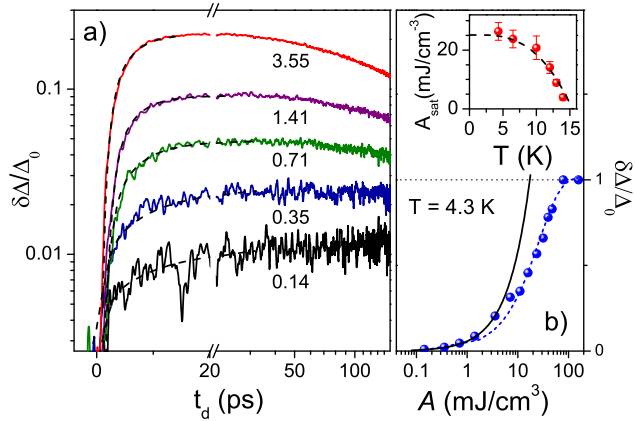}%
\caption{(color online) a) The relative change in gap, $\delta\Delta
/\Delta_{0}$ recorded at 4.3 K for various \emph{A }in mJ/cm$^{3}$. The dashed
lines are fits to the data with Eq.(1). b) The dependence of $\delta
\Delta/\Delta_{0}$ on \emph{A }at 4.3 K. The dashed line is a fit to the
simple saturation model, the solid line is the linear fit. Inset: The
T-dependence of the saturation energy density, $A_{sat}$, compared to the
T-dependence of $\Delta^{2}$ (dashed line).}%
\end{center}
\end{figure}

Fig. 3a) presents the $\delta\Delta(t_{d})$ traces for various \emph{A}
recorded at 4.3 K. As shown in Fig. 3b) the maximum induced change initially
increases linearly with excitation density, followed by a saturation resulting
from suppression of SC. To estimate the characteristic $A$ required to
suppress SC, we use a simple saturation model, $\delta\Delta/\Delta_{0}=$
$\left(  1-\exp\left(  -A/A_{sat}\right)  \right)  $. Here $A_{sat}$ is the
characteristic absorbed energy density required to suppress SC and is
$A_{sat}\left(  4.3\text{ K}\right)  \approx25$ mJ/cm$^{3}$. We find that in
NbN $A_{sat}$ is, within the experimental accuracy, comparable to the
thermodynamic SC condensation energy, $E_{c}=22$ mJ/cm$^{3}$. Here
$E_{c}=B_{c}^{2}/2\mu_{0}=N(0)\Delta^{2}/2$, with $B_{c}=0.234$ T
\cite{Geibel} being the thermodynamic critical field and $N(0)$ the single
spin density of states at the Fermi level. As shown in inset to Fig. 3b), the
T-dependence of $A_{sat}$ is found to follow the T-dependence of $\Delta^{2}$
(dashed line). These observations are in strong contrast to similar studies in
cuprates \cite{Kusar,Pashkin,Beyer}, where $A_{sat}$ is found to be about one
order of magnitude higher than $E_{c}$ and nearly T-independent \cite{Beyer}.

As demonstrated in Fig. 3a), both the CPB and the SC recovery show pronounced
dependence on \emph{A}. The CPB rate is found to increase with increasing
\emph{A}, similarly to the case of MgB$_{2}$ \cite{MgB2}. This behavior can be
attributed to the intrinsic non-linearity of the SC state relaxation process,
where the populations of the photoexcited QP and the high frequency
($\hbar\omega>2\Delta$) phonon (HFP) densities are described by the coupled RT
rate equations \cite{RothwarfTaylor,MgB2,RT}. Here the long timescale of the
CPB implies that photoexcited hot electrons (holes) initially generate a high
density of HFP, which subsequently break Cooper pairs until the
quasi-equilibrium between QP and HFP populations is reached \cite{MgB2,RT}.

The CPB can be quantitatively analyzed using the RT model. In the low
photoexcitation limit ($\delta\Delta<<\Delta$), where the RT analysis is
applicable, $\delta\Delta$ is proportional to the photoexcited QP density,
$n_{PI}$. It is easy to show that at 4K $n_{PI}$ is for the entire range of
\emph{A} substantially higher than the density of thermally excited QP,
$n_{T}$. From $n_{T}\simeq N(0)\sqrt{2\pi\Delta k_{B}T}\exp(-\Delta/T)$, where
$N(0)=0.44$ spin$^{-1}$unit cell$^{-1}$ eV$^{-1}$,\cite{Weber} it follows that
$n_{T}\left(  4\text{ K}\right)  $ $\approx1$ $10^{-6}$ unit cell$^{-1}$ while
$A=0.14$\text{ mJ/cm}$^{3}$\text{ corresponds to }$n_{PI}\approx3$ $10^{-5}$
unit cell$^{-1}$. Therefore, the CPB ($t_{d}\lesssim20$ ps), which is well
separated from the SC state recovery dynamics ($\gtrsim100$ ps), can be fit
(dashed lines in Fig. 3a)) with \cite{MgB2,RT}
\begin{equation}
n_{PI}\left(  t_{d}\right)  =\frac{\beta}{R}\left[  -\frac{1}{4}-\frac
{1}{2\tau}+\frac{1}{\tau}\frac{1}{1-K\exp\left(  -t_{d}\beta/\tau\right)
}\right]  .\label{Eq1}%
\end{equation}
Here $\beta$ is the CPB probability by absorption of HFP, $R$ is the bare QP
recombination rate, while $K$ and $\tau$ are dimensionless parameters (Eq.(3)
of Ref.\cite{MgB2}) determined by $\beta$, $R$ and the initial conditions (the
ratio of the absorbed energy in the HFP and QP subsystems following the
initial e-e and e-ph scattering of hot electrons (holes)), which should be
independent on \emph{A} for these low excitation densities \cite{MgB2,RT}. The
extracted dependences of $\tau/\beta$ and $K$ on \emph{A} are shown in Figs.
4a) and 4b), together with the fit with Eqs.(3) of Ref.\cite{MgB2} (dashed
lines). The best fit is obtained when 91 \% of \emph{A} is initially
transferred to the HFP subsystem, giving the values of the microscopic
constants $\beta^{-1}=6\pm1$ ps, $R=$ $160\pm20$ ps$^{-1}$unit cell.%

\begin{figure}
[ht]
\begin{center}
\includegraphics[
height=6.7151cm,
width=8.4987cm
]%
{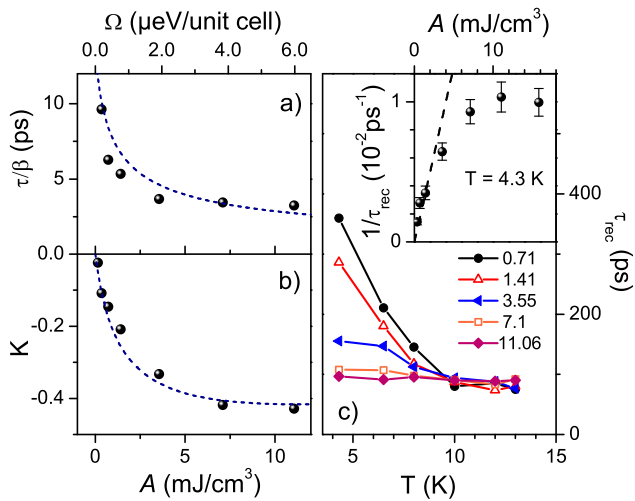}%
\caption{(color online) Panels a) and b) show the dependence of $\tau/\beta$
and $K$ on $A$; the values are extracted from fits to the Cooper pair breaking
dynamics (Fig. 3a). Panel c) shows the T-dependence of the SC recovery time,
$\tau_{rec}$, for several $A$ (in mJ/cm$^{3}$). Inset: $\tau_{rec}^{-1}\left(
A\right)  $ recorded at 4.3 K. For low $A$ the relaxation rate increases
linear with $A$ (dashed line). }%
\end{center}
\end{figure}

Fig. 4c) presents the dependence of the SC recovery time, $\tau_{rec}$, on T
and \emph{A}. Here $\tau_{rec}$ is obtained by fitting the recovery dynamics
with a single exponential decay. As shown in inset to Fig. 4c), at 4 K
$\tau_{rec}^{-1}$ first increases linearly with \emph{A}, mimicking the
intrinsic bimolecular kinetics of the QP recombination \cite{RT}. For
$\emph{A}\gtrsim5$ mJ/cm$^{3}$ and for $T>10$ K $\tau_{rec}^{-1}$ is constant
at $\tau_{rec}^{-1}(0)\approx0.01$ ps$^{-1}$. In NbN $\tau_{rec}^{-1}(0)$ is
governed by the escape of the HFP into the substrate and is inversely
proportional to the film thickness \cite{Ilin}. It was argued that the
dependence of $\tau_{rec}$ on \emph{A}, observed in cuprates, implies the
absence of the phonon bottleneck in cuprates \cite{Gedik,KaindlBiSCO}. The
subsequent analysis of the RT model \cite{RT}, however, suggested a (perhaps
counterintuitive) result, that also in a phonon bottleneck case $\tau
_{rec}^{-1}\propto n_{PI}$ for $n_{T}<n_{PI}\leq\beta/R$. Here $\beta/R$ is
the material dependent characteristic QP density, below which the RT
perturbative description is applicable \cite{RT}. Indeed, $A=5$ mJ/cm$^{3}$
corresponds to $n_{PI}\approx0.001$ unit cell$^{-1}$, identical to
$\beta/R\approx0.001$ unit cell$^{-1}$ determined from the analysis of the
CPB. Similarly, the fact that $\tau_{rec}^{-1}$ is above $\approx10$ K $A$-
and T-independent follows from the fact that $n_{T}(10K)\approx0.0003$ unit
cell$^{-1}$ is comparable to $\beta/R$. Therefore, as experimentally
demonstrated here for a standard BCS SC, the \emph{A}-dependent SC state
recovery at low Ts seems to be an intrinsic feature of a SC driven out of
equilibrium, and should be observed in the phonon bottleneck case, providing
that $n_{T}<n_{PI}\leq\beta/R$. Based on this study, we argue that it is the
absence of \emph{A}-dependence of the SC state recovery, observed e.g. in
optimally doped YBCO \cite{Kabanov} and in an overdoped BSCCO \cite{Gedik},
that presents an anomalous behavior and not vice versa as argued
\cite{Gedik,KaindlBiSCO}.

As demonstrated, the dynamics of photoexcited NbN can be over
large range of T and \emph{A} well described by the phenomenological RT model,
enabling us to determine the values of microscopic parameters $R$ and $\beta$.
Since $R=\frac{8\pi\lambda\Delta^{2}}{\hbar N\left(  0\right)  \Omega_{c}^{2}%
}$ \cite{mercury,Ovchinikov}, where $\Omega_{c}$ is the phonon cut-off
frequency, the value of $\lambda$ can be determined. Taking the known values
for $\Delta$, $N(0)$ and\ $\Omega_{c}=16$ THz \cite{Geibel}, we obtain
$\lambda=1.1\pm0.12$, which is in a very good agreement with the theoretical
estimates, $\lambda=1-1.12$ \cite{Weber,lambda}. Importantly, such an approach
does not suffer from the (ambiguous) underlying assumptions of the
two-temperature model, the model commonly \cite{Perfetti,Mansart} used to
determine $\lambda$.

In summary, we presented the first systematic studies of the SC state
relaxation phenomena in a conventional BCS superconductor NbN with ps
time-resolution. Utilizing the TDTS, we were able to study the time evolution
of $\Delta$ over large range of temperatures and excitation densities. We
demonstrated, that both the CPB and SC state recovery dynamics depend strongly
on the excitation density in agreement with the predictions of the
phenomenological RT model \cite{MgB2,RT}. Studying the CPB enabled us to determine the values
of the microscopic parameters $R$, $\beta$, as well as the dimensionless e-ph
coupling constant $\lambda$ in NbN. This approach could be used to determine
the electron--boson coupling strengths in high-T$_{c}$ SCs, providing the CPB
dynamics as a function of excitation density can be experimentally resolved. Last but not least, our results on NbN emphasize the
unusual nature of the \emph{A}-independent relaxation observed in some cuprates
\cite{Kabanov,Gedik}.

This work was supported by the German Israeli DIP project No. 563363,
Alexander von Humboldt Foundation, Zukunftskolleg
and Center for Applied Photonics at the University of Konstanz.

\end{document}